\documentclass[prl,twocolumn,showpacs,superscriptaddress]{revtex4}                    

\usepackage{graphicx,amssymb,amsmath,color,bbm,psfrag}
\usepackage[normalem]{ulem}

\begin{document}


\newcommand{\dirop}{\mathcal{D}}

\newcommand{\cro}[1]{\textcolor{red}{}}
\newcommand{\edi}[1]{{#1}}

\title{Diverging dc conductivity due to a flat band in disordered pseudospin-1 Dirac-Weyl fermions} 
\author{M\'at\'e Vigh}
\affiliation{Department of Physics of Complex Systems, E\"otv\"os University, H-1117 Budapest, Hungary}
\author{L\'aszl\'o Oroszl\'any}
\affiliation{Department of Theoretical Physics, Budapest University of Technology and
  Economics, Budapest, Hungary}
\author{Szabolcs Vajna}
\affiliation{Department of Physics and BME-MTA Exotic  Quantum  Phases Research Group, Budapest University of Technology and
  Economics, Budapest, Hungary}
\author{Pablo San-Jose}
\affiliation{Instituto de Ciencia de Materiales de Madrid (ICMM-CSIC), Cantoblanco, 28049 Madrid, Spain}
\author{Gyula D\'avid}
\author{J\'ozsef Cserti}
\affiliation{Department of Physics of Complex Systems, E\"otv\"os University, H-1117 Budapest, Hungary}
\author{Bal\'azs D\'ora}
\email{dora@eik.bme.hu}
\affiliation{Department of Physics and BME-MTA Exotic  Quantum  Phases Research Group, Budapest University of Technology and
  Economics, Budapest, Hungary}

\date{\today}

\begin{abstract}
Several lattices, such as the dice or the Lieb lattice,  possess Dirac cones and a flat band 
crossing the Dirac point, whose effective model is the pseudospin-1 Dirac-Weyl equation.
 We investigate the fate of the flat band in the presence of disorder
by focusing on the density of states (DOS) and dc conductivity.
While the central hub-site does not reveal the presence of the flat band, 
the sublattice resolved DOS on the non-central sites exhibits a narrow  peak with height $\sim 1/\sqrt{g}$ with $g$ the dimensionless disorder variance.
Although the group velocity is zero on the flat band, the dc conductivity diverges as $\ln(1/g)$ with decreasing disorder
due to interband transitions around the band touching point between the propagating and the flat band.
Generalizations to higher pseudospin are given.
\end{abstract}


\maketitle

\emph{Introduction.} 
Flat bands are at the heart of several peculiar phenomena in condensed matter, especially in the presence of strong correlations. Prominent examples include (nearly-) flat band
ferromagnetism, integer and fractional quantum Hall effect arising from Landau levels in finite magnetic field, and recently fractional quantum Hall effect at zero magnetic field\cite{flat1,flat2,flat3}.
Engineering flat bands with topologically non-trivial character has become a major challenge recently in connections with topological insulators\cite{volovik,hasankane}.

\begin{figure}[h!]
\psfrag{A}[][][1][0]{\color{magenta}$A$}
\psfrag{H}[][][1][0]{\color{blue}$H$}
\psfrag{B}[][][1][0]{$B$}
\includegraphics[width=3cm]{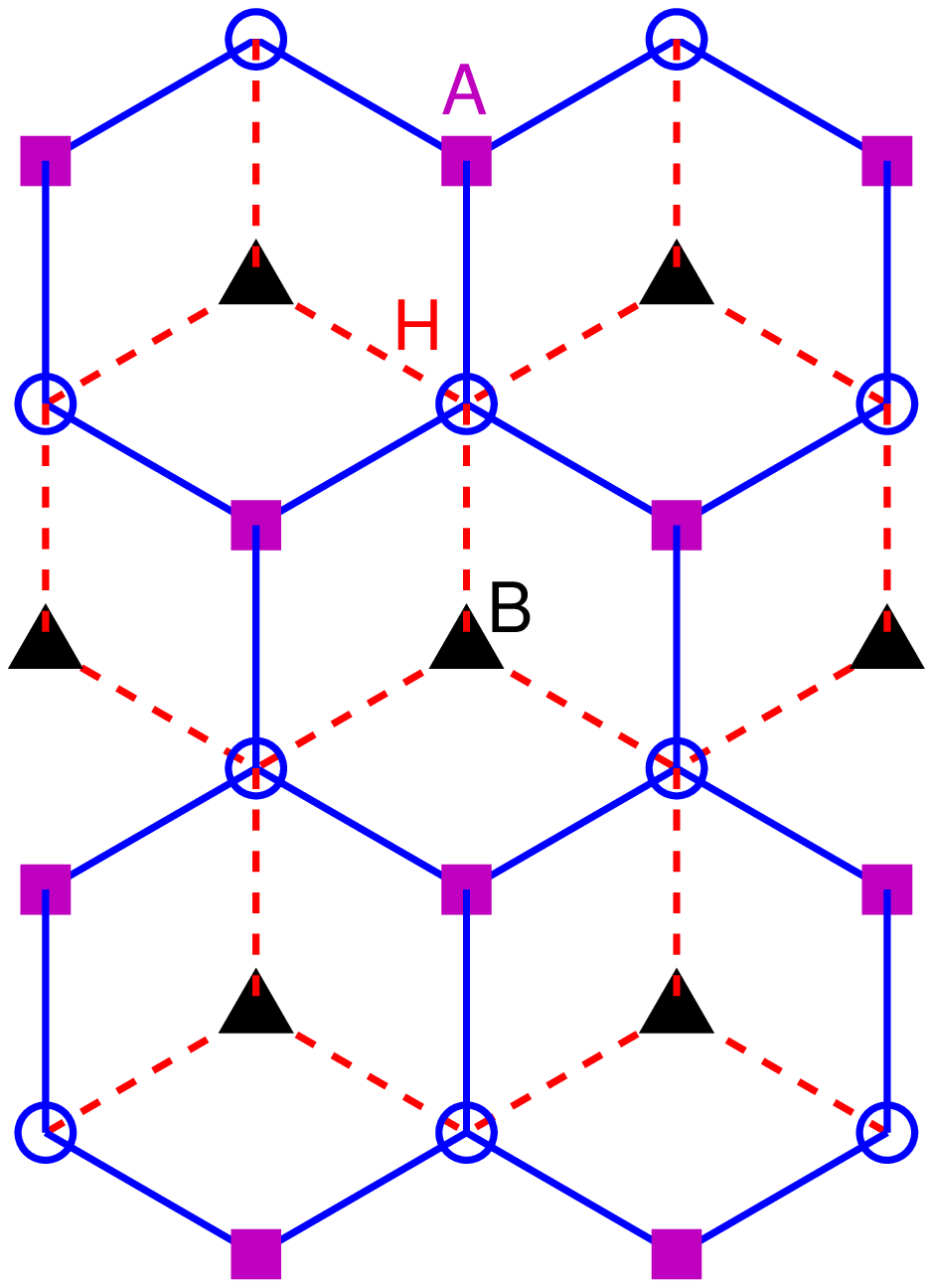}
\psfrag{x}[b][t][1][0]{$p_x$}
\psfrag{y}[b][t][1][0]{$p_y$}
\psfrag{z}[b][t][1][0]{$E({\bf p})$}
\includegraphics[width=4cm]{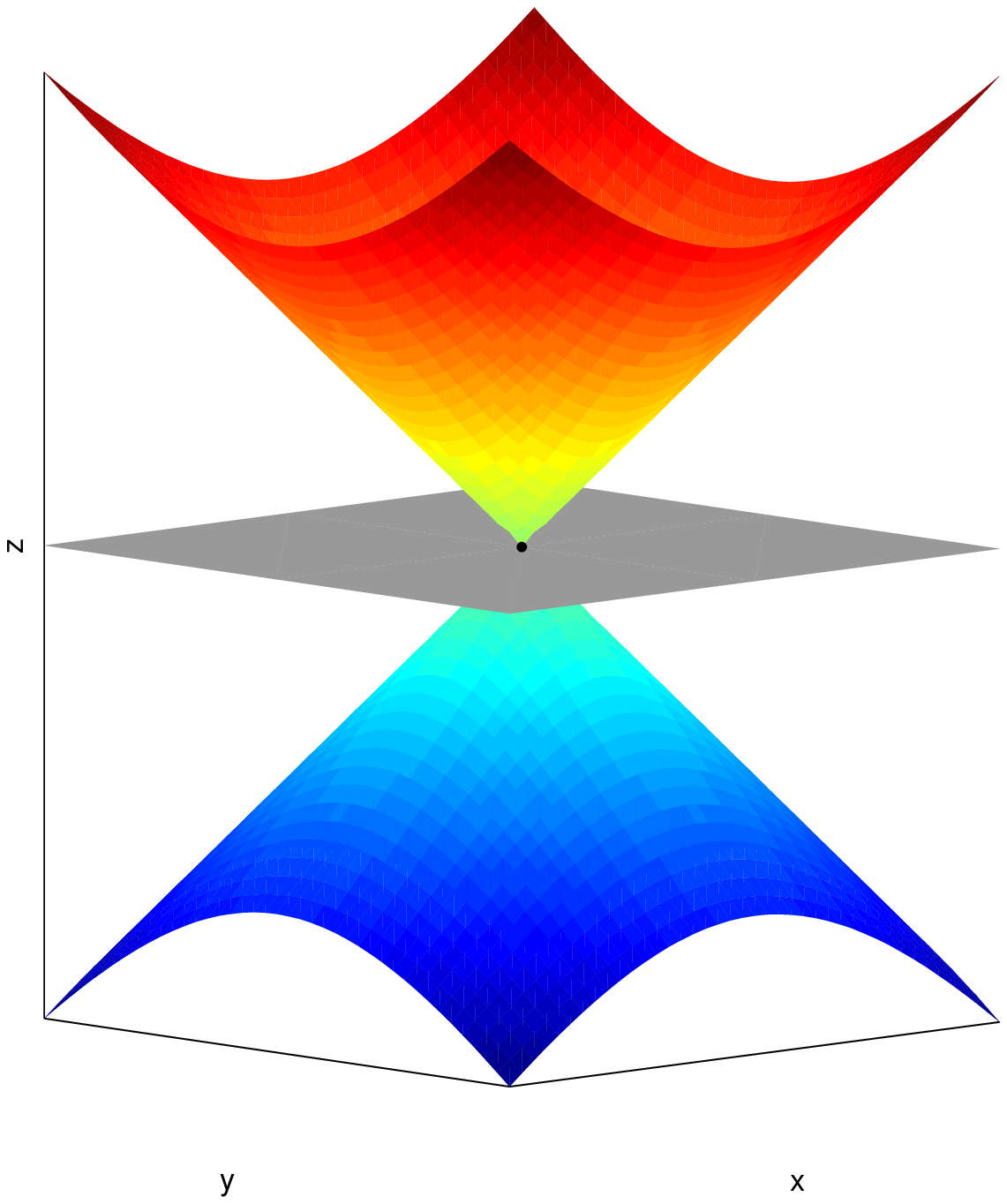}
\caption{A small segment of the dice or $T_3$ lattice, consisting of two, one sublattice sharing (circles, hub site) honeycomb lattices, 
is shown (left), where the lines denote uniform hoppings $t$, together with its low energy
spectrum around the corners of the hexagonal Brillouin zone, featuring a flat band at the touching point of the Dirac cones (right).}
\label{t3lattice}
\end{figure}

In addition to flat bands as surface modes, these also appear as bulk bands in systems with specific two-dimensional lattice structures, such as the dice or $T_3$ lattices, the Lieb lattice etc.
The common feature in the  spectrum of these lattices is a graphene-like Dirac cone, intersected by a completely dispersionless flat band at the Dirac point (see Fig. \ref{t3lattice}).
These can be regarded as the pseudospin-1 generalization of the Dirac equation \cite{bercioux,shen,greens1,apaja,urban}, and arise in the family of 
higher pseudospin generalizations of the Dirac equation,  proposed in Refs. \cite{lan,watanabe,diracdora}. 

While many of their properties are well understood, including topology, not much is known about their transport properties\cite{urban}, which promise many excitement in light
of the fascinating transport properties of their pseudospin-1/2 counterpart in graphene. There, the universal value of the minimal conductivity at half filling \cite{peresrmp} 
attracted significant attention over the years, whereby the  decreasing number of charge carriers as the charge neutrality point is approached exactly compensates their increasingly long lifetime. 

Charge transport in the pseudospin-1 family of Dirac-Weyl \cro{equations} \edi{fermions} seems to be non-trivial as well. Due to the flat-band, the density of states (DOS) exhibits a sharp peak
at the neutrality point, though the group velocity on the flat band is identically zero. Consequently, at least two scenarios seem plausible  for the behaviour of the dc conductivity: it can  remain insensitive to the flat band or 
it can be influenced by the large number of available states on the flat band. 
Intriguingly,  none of these \edi{simple} pictures are {completely} correct: while the dc conductivity solely from the flat band vanishes 
due its zero velocity, interband transition between the flat band and adjacent propagating bands
are possible at the band touching degeneracy point, and this transition causes the divergence of the dc 
conductivity  in the pure system precisely  due to the zero flat band velocity. 
In the presence of disorder, this divergence is cut-off logarithmically by the disorder strength, as we show by a careful numerical and analytical investigation of the problem.

Experimentally, the dice lattice can be realized in  SrTiO$_3$/SrIrO$_3$/SrTiO$_3$ trilayer heterostructures\cite{wangran}, and the observation of the predicted
features may be realized. Additionally, the dice or Lieb lattices can be created  via optical means in the cold atomic setting, and disorder can be introduced in a controlled manner using
 speckle potentials\cite{blochrmp}. While the local DOS has long been measurable by e.g. time of flight imaging or rf spectroscopy\cite{blochrmp}, the dc conductivity is also accessible in this setting as well\cite{Brantut}.

\emph{Density of states.} 
The dice lattice, shown in Fig. \ref{t3lattice}, consists of a sixfold coordinated hub-site ($H$), and two threefold coordinated rim-sites ($A$ and $B$) within its unit cell, 
with uniform hopping integrals $t$.
Its Brillouin zone is hexagonal, and contains
 low energy \edi{excitations} close to zero energy at the two non-equivalent corners of the Brillouin zone\cite{bercioux}, similarly to graphene\cite{castro07}. These are described by the
pseudospin-1 Dirac-Weyl equation  as\cite{diracdora}
\begin{gather}
H_0=v_F\bf{S\cdot p},
\label{ham}
\end{gather}
where $v_F$ is the Fermi velocity, ${\bf p}=(p_x,p_y)$ and
\begin{gather}
S_x=\frac{1}{\sqrt 2}\left(
\begin{array}{ccc}
0&  1& 0\\
1& 0& 1\\
0& 1& 0
\end{array}
\right),\hspace*{3mm}
S_y=\frac{1}{\sqrt 2}\left(
\begin{array}{ccc}
0&  -i& 0\\
i& 0& -i\\
0& i& 0
\end{array}
\right).
\end{gather}
The resulting band-structure consists of three bands as $E_{\pm}({\bf p})=\pm v_F |p|$ and 
 $E_{0}({\bf p})=0$.
The density of states (DOS) reads as 
\begin{gather}
\rho(\omega)= \frac{A_c}{2\pi}\frac{|\omega|}{v_F^2} +\delta(\omega)=\frac{2|\omega|}{D^2} +\delta(\omega)
\label{dos}
\end{gather}
for $|\omega|<D$ per spin, valley and unit cell, $A_c=4\pi/k_c^2$ being the unit cell area. 
\edi{Here,} $k_c$ is the momentum space cutoff and $D=v_Fk_c$ is the half-bandwidth. The DOS satisfies $\int_{-D}^Dd\omega \rho(\omega)=3$ and remains
linear in energy close to half filling, similar to graphene, but exhibits a sharp peak due to the  
flat band\cite{bercioux} at zero energy.

The effect of weak disorder is modeled  by adding a short range Gaussian potential as 
\begin{gather}
U({\bf r})=\left(
\begin{array}{ccc}
U_A({\bf r})  &0 & 0\\
0& U_H({\bf r}) & 0\\
0&0& U_B({\bf r})
\end{array}
\right),
\label{impurityham}
\end{gather}
where $\overline{ U_A({\bf r})U_A({\bf r}^\prime)}=U^2\delta({\bf r-r}^\prime)$ \edi{(and similarly for sublattice $H$ and $B$)} with no intersublattice disorder correlation\edi{. The} overline
represents disorder averaging.
To determine the structure of the self-energy, we study the effect of disorder within the self-consistent Born approximation (SCBA)
as
$\Sigma(i\omega_n)={U^2}{}\sum_{\bf p}G({\bf p},i\omega_n)/N$,
where $N$ is the number of unit cells\edi{,}
$G({\bf p},i\omega_n)=\left(i\omega_n-H_0-\Sigma(i\omega_n)\right)^{-1}$\edi{,}
\edi{and $\omega_n$ is the fermionic Matsubara frequency.} The off-diagonal elements of the Green's function vanish after momentum integration, and the self-energy
reads as
\begin{gather}
\Sigma(i\omega_n)=\left(
\begin{array}{ccc}
\Sigma_A(i\omega_n) &0 &0\\
0& \Sigma_H(i\omega_n) & 0\\
0 & 0& \Sigma_A(i\omega_n)
\end{array}
\right)\edi{.}
\end{gather}
 The self-energy highlights the distinct structure of sublattice $H$, i.e. hopping from $A$ to $B$ is only possible through $H$.
The self-consistency equations are expressed as
\begin{subequations}
\begin{gather}
\Sigma_A(i\omega_n)=\frac{g}{8}\left(\frac{D^2}{z_1}-z_0\ln\left(1-\frac{D^2}{z_0z_1}\right)\right),\\
\Sigma_H(i\omega_n)=-g\frac{z_1}{4}\ln\left(1-\frac{D^2}{z_0z_1}\right),
\end{gather}
\label{sigmas}
\end{subequations}
where $g=U^2A_c/\pi v_F^2=4U^2/D^2$ is the dimensionless disorder strength, and $z_{0,1}=i\omega_n-\Sigma_{H,A}(i\omega_n)$. 
\edi{Long range disorder corresponds to $U_A({\bf r})=U_H({\bf r})=U_B({\bf r})$ in Eq. (\ref{impurityham}), which interestingly yields the same self-energy. 
Similarly, randomly} distributed substitutional impurities with strength $U_{i}$ and concentration $n_i$
also give the same self-energy in the Born approximation with $g\sim n_iU_{i}^2$, though with different  numerical prefactors.

\begin{figure}[h!]
\psfrag{x}[t][b][1][0]{$U/ t, 2 \sqrt{g}$}
\psfrag{y}[b][t][1][0]{ {\color{red}$\rho_H(0)t 2\pi^2$}, {\color{blue}$1/\rho_A(0)t$}}
\psfrag{xx}[t][b][1][0]{$g$}
\psfrag{yy}[b][t][1][0]{$\rho(0) D$}
{\includegraphics[width=5cm]{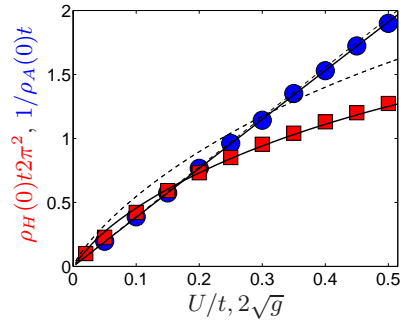}}
\caption{The zero energy  DOS of the $T_3$ lattice is shown from KPM, the
red squares/blue circles corresponding to sublattice resolved DOS of $H$ and the inverse DOS of $A$, respectively. The solid
lines are the results from the SCBA using Eqs. \eqref{sigmas} and \eqref{dosscba} (with an overall normalization factor as fitting parameter), 
while the dashed lines stem from the approximate expressions in Eq. \eqref{zerodos}.
\label{doss1}}
\end{figure}

 In general, these equations can only be solved numerically by e.g. iteration, 
but analytical
treatment is possible in certain limiting cases. At the Dirac point,
after analytical continuation to real frequencies, 
we obtain
$\Sigma_A(0)=-i\sqrt{g} {D}/{2^{3/2}}$, 
$\Sigma_H(0)\simeq-i g^{3/2}{D}\ln\left({32}/{g^2}\right)/8\sqrt 2$. 
The self-consistency equation of $\Sigma_H(0)$ in the \emph{Born} 
limit parallels closely the self energy of graphene\cite{ostrovsky} and d-wave superconductors\cite{impurd-wave} in the \emph{unitary} limit.

The knowledge of the self-energies gives immediate access to the sublattice resolved DOS, which
reads after analytical continuation as
\begin{gather}
\rho_{A/H}(\omega)=- \frac{\textmd{Im}\left[\Sigma_{A/H}(i\omega_n\rightarrow \omega+i\delta)\right]}{\pi U^2}
\label{dosscba}
\end{gather}
per spin and valley, respectively, $\delta\rightarrow 0^+$\edi{. The} DOS on sublattice $B$ is identical to $A$.
The total DOS is then $2\rho_A(\omega)+\rho_H(\omega)$, which is dominated by the first term at low energies, 
the second one only contributes to the linear in energy region for $ D\gg |\omega|\gg \sqrt{g/2}D$.

\begin{figure}[h!]
\psfrag{x}[t][b][1][0]{$\omega/t$}
\psfrag{y}[b][t][1][0]{$\rho(\omega) t$}
\psfrag{xx}[t][b][1][0]{$\omega/D$}
\psfrag{yy}[b][t][1][0]{$\rho(\omega)D$}
\psfrag{KPM, sublattice H}[][][0.9][0]{KPM, sublattice H}
\psfrag{KPM, sublattice A}[][][0.9][0]{KPM, sublattice A, B}
\psfrag{SCBA, sublattice H}[][][0.9][0]{SCBA, sublattice H}
\psfrag{SCBA, sublattice A}[][][0.9][0]{SCBA, sublattice A, B}
{\includegraphics[width=8cm]{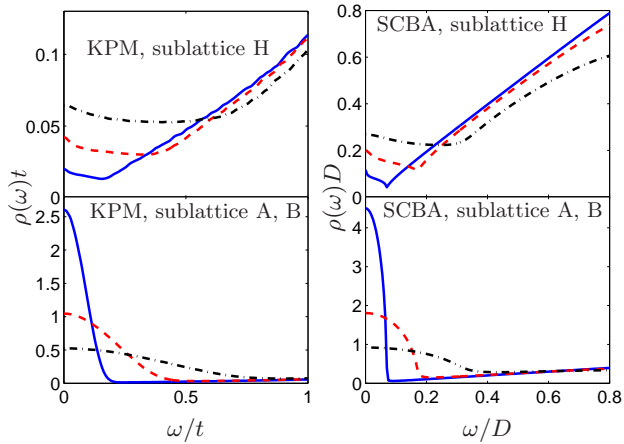}}
\caption{The sublattice resolved DOS is shown from the KPM calculation on the $T_3$ lattice for Gaussian disorder with standard deviation $U/t$=0.1 (blue solid), 0.25 (red rashed) and 0.5 (black dash-dotted) in left panel.
The right panel visualizes the SCBA results for $\sqrt g=0.1$ (blue solid), 0.25 (red rashed) and 0.5 (black dash-dotted).\label{dost3}}
\end{figure}

This reproduces the linear in energy DOS at high energies and weak disorder. Since the 
flat band lives on sublattice $A$ and $B$\cite{diracdora} (i.e. its spinor component on $H$ is zero), 
the local DOS on sublattice $H$ does not probe its presence in the spectrum, therefore
it does not contain the Dirac-delta peak, only the graphene like linear in energy part.  
In the presence of disorder, the zero energy peak \edi{in $A$ and $B$ sublattices} broadens with width $\sqrt{g/2} D$, \edi{while}  $\rho_H(\omega\simeq 0)$ develops a flat plateau\edi{. Their} heights are 
\begin{gather}
\rho_A(0)=\frac{\sqrt 2}{\pi D \sqrt g}, \hspace*{5mm} \rho_H(0)=\frac{\sqrt g}{\pi D 2^{3/2}} \ln\left(\frac{32}{g^2}\right).
\label{zerodos}
\end{gather}
In spite of the fact that the DOS on sublattice $H$ is graphene like for the clean system, the effect of disorder on its low energy part is completely distinct from that 
in graphene\cite{ando1,ostrovsky}.
These features are shown in Figs. \ref{doss1} and \ref{dost3}, together with the numerical results on the dice lattice as in Fig. \ref{t3lattice}, containing 2000x2000 sites 
 with Gaussian distributed
potential disorder (with zero mean  and $U^2$ variance) 
using the 
kernel polynomial method (KPM)\cite{kpm}. The SCBA for Eq. \eqref{ham} in the continuum limit reproduces all the main features seen in the numerics for the dice lattice, 
including a dip in the H sublattice DOS at $\sqrt{g/2} D$.

\emph{dc conductivity.}
 The dc conductivity is evaluated after dressing the current-current correlation function with impurity lines. The current operator
in the $x$ direction is $j_x=ev_FS_x$. Assuming short range scatterers, only self-energy corrections are present, vertex corrections 
vanish, similarly to the case of graphene\cite{ostrovsky,ando1}.
Long range scatterers would yield finite vertex corrections, though.
Then, the dc conductivity is obtained from the Kubo formula  at a given chemical potential, $\mu$\edi{. At} zero temperature  
after some algebraic manipulation \edi{one obtains}
\begin{gather}
\sigma_{xx}=\frac{1 }{\pi A}\sum_{\bf p}\textmd{Tr}\left[G({\bf p},\mu+i\delta)j_xG({\bf p},\mu-i\delta)j_x\right]
\end{gather}
per spin and valley, $A=NA_c$ is the total area of the system and $\sigma_{yy} = \sigma_{xx}$ and the off-diagonal elements are zero.
Performing the momentum integral, and using Eqs. \eqref{sigmas},
we finally get, upon restoring original units, 
\begin{gather}
\sigma_{xx}=2\sigma_0\textmd{Re}\left[\frac{2i|x_1^2|\textmd{Re}(x_0)+x_0\textmd{Im}(x_1^2)}{|x_1^2|\textmd{Im}(x_0x_1)g}\Sigma_H(\mu)\right],
\label{sigmageneral}
\end{gather}
where $x_{0,1}=\mu-\Sigma_{H,A}(\mu)$, and $\sigma_0=e^2/\pi h$ is the universal minimal conductivity \edi{per spin and valley} of pseudospin-$\frac 12$ Dirac fermions\cite{csertinote} at half filling, also found by numerical studies\cite{nomura}.

\begin{figure}[h!]
\psfrag{xx}[t][b][1][0]{$\mu/ D$}
\psfrag{yy}[b][t][1][0]{$\sigma_{xx}(\mu)3g/8\sigma_0$}
\psfrag{x}[t][b][1][0]{$\mu/t$}
\psfrag{y}[b][][1][0]{$\sigma_{xx}(\mu)3U^2/16 t^2\sigma_0$}
\psfrag{KPM}[][][1][0]{KPM}
\psfrag{SCBA}[][][1][0]{SCBA}
{\includegraphics[width=8cm]{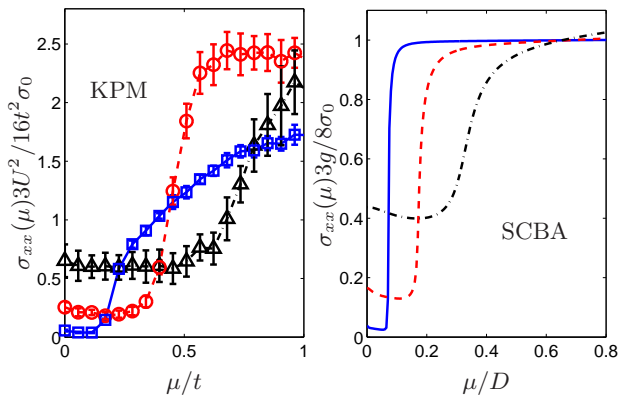}}
\caption{The scaled dc conductivity per spin and valley for $U/t=0.1$ (blue solid), 0.25 (red dashed) and 0.5 (black dash-dotted) 
is shown as a function of the chemical potential from the KPM calculation of the $T_3$ lattice with 2000x2000 sites (left panel). The right panel visualizes the
the same quantity calculated from SCBA in the continuum limit for $\sqrt g=0.1$  $0.25$ and $0.5$ with the same color coding.\label{dccondmu}}
\end{figure}

For finite doping, after expanding the self energy to second order in $g$ using Eqs. \eqref{sigmas}, and plugging the resulting expressions \edi{into} Eq. \eqref{sigmageneral},
we obtain the conductivity. It becomes practically constant  with a weak logarithmic doping dependence for $\sqrt{g/2}\ll |\mu|/D\ll 1$ as
\begin{gather}
\sigma_{xx}=\sigma_0\frac{8}{3g}\left(1-\frac{g}{4}\ln\left(\frac{D}{|\mu|}\right)\right),
\end{gather}
which agrees qualitatively with that of graphene\cite{ostrovsky}. Due to the finite doping, this is dominated by intraband contributions, and the presence of the flat
band does not play a role. The weak chemical potential dependence of the conductivity \edi{is} plotted in Fig. \ref{dccondmu}, together with the numerical computation of the Kubo formula using
KPM\cite{ferreira}.

\begin{figure}[h!]
\psfrag{x}[t][b][1][0]{$(U/t)^2$, $4g$}
\psfrag{y}[b][t][1][0]{$\sigma_{xx}(0)/\sigma_0$}
\includegraphics[width=5.5cm]{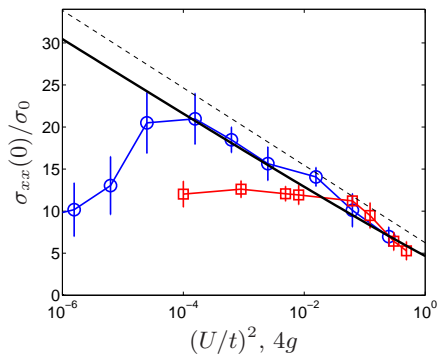}
\caption{The dc conductivity per cone is shown as a function of the disorder variance from the KPM on the dice lattice for 2000x2000 (blue circles) and 600x600 (red squares) sizes, errorbars indicated, solid lines
connecting the symbols are guides to the eye. The SCBA yields the solid black line from the first equality in Eq. \eqref{dccond}, while the thin black dashed line is the approximate $\ln(32/g^2)$ expression.}
\label{dccondnumer}
\end{figure}

At the charge neutrality point,
Eq. \eqref{sigmageneral} simplifies to
\begin{gather}
\frac{\sigma_{xx}}{\sigma_0}=\ln\left(1-\frac{2D^2}{\Sigma_H(0)\Sigma_A(0)}\right)=\frac{4\Sigma_H(0)}{\Sigma_A(0)g}\simeq  \ln\left(\frac{32}{g^2}\right),
\label{dccond}
\end{gather}
and remains
roughly unchanged for finite doping in the $|\mu|<\sqrt{g/2}D$ range, as seen in Fig. \ref{dccondmu}.
The agreement between KPM and SCBA becomes better with increasing system sizes: the average level spacing in the propagating bands is $\sim t/N$, while the broadening of the flat band is $\sim U$.
As long as the former is smaller than the latter, the numerics reproduces Eq. \eqref{dccond}.
Therefore, the dc conductivity diverges with decreasing disorder strength as $\ln(1/g)$ due to the presence of interband transitions between adjacent bands:
when bands touch (flat and propagating, see Fig. \ref{t3lattice}) at the Dirac point, interband transitions are possible in the dc  limit.

This can be understood in the constant relaxation time approximation, when an interband transition 
contributes to $\sigma_{xx}$ at the Dirac point with $\int d^2k |M_{n,n+1}|^2\rho_n(k)\rho_{n+1}(k)$, where
$M_{n,n+1}$ is the matrix element between band $n$ and $n+1$\edi{. Here,} $\rho_n(k)=\Gamma/(\Gamma^2+\varepsilon_n^2(k))\pi$ and $\varepsilon_n(k)=v_n|k|$ are
 the spectral function and energy dispersion of the $n$th band, respectively, \edi{and 
$\Gamma$ is} the scattering rate.
In the case of two linearly dispersing bands, this yields  
\edi{$\sigma_{xx}\sim\ln(|v_n/v_{n+1}|)$}.
In the case of a flat band, $v_n=0$ and the integral is both infrared and ultraviolet divergent for $\Gamma=0$ due to its dispersionless spectrum. 
While the latter is cured by a natural high energy cutoff $D$, 
the former requires another energy scale, i.e. the scattering rate, yielding the above log-divergence in Eq. 
\eqref{dccond}, as seen in Fig. \ref{dccondnumer}, confirmed also numerically from the Kubo formula using
KPM.
Such transitions are in principle also present for graphene at the Dirac point, but vanish due to the identical velocities in the upper and lower Dirac cones.

For higher pseudospin generalizations of the Dirac equation\cite{diracdora,lan}, the
interband transitions between two adjacent, propagating bands give a finite contribution the 
dc conductivity
at the Dirac point. For a perfect pseudospin-$S$ Dirac equation with half-integer $S$, the spectrum is $E_{n}({\bf p})=nv_F|{\bf p}|$ with $n=-S\dots S$, thus 
all bands are propagating. The universal minimal conductivity is
\begin{gather}
\sigma_S(\mu=0)=\frac{\sigma_0}{2}\left(\left(S+\frac 12\right)\left(S+\frac 32\right)+\right.\nonumber\\
\left.+\sum\limits_{n=1/2}^S\frac{S(S+1)-n(n+1)}{2n+1}\ln\left(\frac{n+1}{n}\right)\right),
\end{gather}
where first term stems from intraband processes and the the log-terms come from interband transitions between adjacent bands with velocities $nv_F$ and $(n+1)v_F$.
As an example, this gives $\sigma_{3/2}=\sigma_0\left(3+\frac 34 \ln(3)\right)$ for $S=3/2$ and grows with $S^2$ for $S\rightarrow \infty$.
For integer $S$, the conductivity diverges due to the flat band as $\sim S(S+1)\ln(1/g)$, similarly to the $S=1$ case in Eq. \eqref{dccond}.

In conclusion, we have studied the disordered $S=1$ Dirac-Weyl equation analytically using the SCBA and numerically by the KPM on the dice lattice. The contribution of the flat band 
is only present in the sublattice resolved DOS for the rim sites. The dc conductivity diverges logarithmically with decreasing disorder due to interband transitions at the band touching point between the flat and
propagating bands from the SCBA, showing excellent agreement with KPM. The divergence of the dc conductivity is a general feature for  integer pseudospin-S Dirac-Weyl fermions, 
and is expected to hold true whenever a propagating and a flat bands touch, i.e. also for the kagome lattice.
Their half-integer pseudospin-S counterpart with no flat band possesses a universal minimal conductivity at half filling, though with a value different from that in graphene.

\begin{acknowledgments}

Supported by the Hungarian Scientific Research Fund
Nos. K101244, 75529 and 81492, CNK80991,
New Sz\'echenyi Plan Nr.
   T\'{A}MOP-4.2.1/B-09/1/KMR-2010-0002\edi{, the European Research Council Advanced Grant, contract 290846,} and ERC Grant Nr. ERC-259374-Sylo
and    by the Marie Curie ITN project NanoCTM (FP7-PEOPLE-ITN-2008-234970).
\end{acknowledgments}

\bibliographystyle{apsrev}

\bibliography{refgraph}

\end{document}